# The relationship between usage and citations in an open access mega journal

Barbara McGillivray

The Alan Turing Institute, London, UK
Theoretical and Applied Linguistics, Faculty of Modern and Medieval Languages, University of Cambridge, Cambridge, UK

Mathias Astell

Hindawi Limited, London, UK

## Abstract

How do the level of usage of an article, the timeframe of its usage and its subject area relate to the number of citations it accrues? This paper aims to answer this question through an observational study of usage and citation data collected about the multidisciplinary, open access mega-journal *Scientific Reports*. This observational study answers these questions using the following methods: an overlap analysis of most read and top-cited articles; Spearman correlation tests between total citation counts over two years and usage over various timeframes; a comparison of first months of citation for most read and all articles; a Wilcoxon test on the distribution of total citations of early cited articles and the distribution of total citations of all other articles. All analyses were performed using the programming language R. As *Scientific Reports* is a multidisciplinary journal covering all natural and clinical sciences, we also looked at the differences across subjects. We found a moderate correlation between usage in the first year and citations in the first two years since publication (Spearman correlation coefficient of 0.49, α=0.05), and that articles with high usage in the first 6 months are more likely to have their first citation earlier (Wilcoxon=1811500, p < 0.0001), which is also related to higher citations in the first two years (Wilcoxon=8071200, p < 0.0001). As this final assertion is inferred based on the results of the other elements of this paper, it requires further analysis. Moreover, our choice of a 2 year window for our analysis did not consider the articles' citation half-life, and our use of *Scientific Reports* (a journal that is atypical compared to most academic journals) as the source of the articles analysed has likely played a role in our findings, and so analysing a longer timeframe and carrying out similar analysis on a different journal (or group of journals) may lead to different conclusions.

## Introduction

Understanding the impact of published research is of utmost importance to the success and continued development of the scientific dialogue. Without knowing if a published research article (or group of articles) has been impactful, it becomes difficult to best direct the development of science (Taylor 2011; Hicks 2011; Grgić 2015). While many metrics have been developed to help understand the impact of published research articles (Bollen et al. 2009), the main focus has historically been on the number of citations articles receive – with the use of citations for this purpose dating as far back as 1927 (Gross 1927). Beyond citations though, the other major metric that is used to understand the impact of published articles is the level of usage they receive (also variously referred to as 'views', 'reads' or 'downloads'). Usage data are most often primarily provided by the publisher of the research article. With the move towards online article publication, data on article usage (online usage at least) has become much more accessible and so more readily available to use in assessing impact (Duy et al. 2006; Armbruster 2010). Usage had previously been taken into account when deciding on the reach or impact of an article, but before journals became available online (either wholly, or alongside a continuing print version), this was based only on the physical circulation of a journal (Peritz 1995; Buffardi & Nichols 1981). The use of print circulation however only provided an indication of usage, as it did not specifically track individual uses of article. The move to effectively tracking online usage has provided a much greater opportunity to understand and assess article impact through usage.

Although citations and usage are two of the primary metrics used to assess impact, an agreed understanding of how each helps in presenting the impact of published articles has not yet been fully reached. While there has been a number of studies and deliberations on the relationship between these two metrics, these have not produced a definitive answer (Perneger 2004; Bollen et al. 2005; Moed 2005; Brody et al. 2006; McDonald 2007; Chu & Krichel 2007; Garfield 2011, Guerrero-Bote & Moya-Anegón 2014). Many of the studies carried out on this topic are also restricted to assessing specific subject areas (Schloegl & Gorraiz 2010; Nieder et al. 2013; Gorraiz 2014) or geographical regions (Vaughan et al. 2017; Chi & Glänzel 2017; Wical & Vandenbark 2015) and even in these cases the results of different studies do not often provide a homogenous indication as to the relationship between citations and usage. Furthermore, although there has been research into the role open access plays on this relationship (Antelman 2004; Davis et al. 2008; Gargouri et al. 2010; Davis 2011; Davis & Walters 2011; Wang et al. 2015) and one study looked at the relationship between tweets and citations and usage (de Winter 2015), there have not been (as far as we know) any studies utilizing data from an open access 'mega-journal' to assess the relationship between usage and citations on a large scale and across multiple subject areas. We also did not find any systematic or literature reviews that covered the scope of our study, and so we were unable to include these in our assessment of previous research. We also recognize that it has been variously argued that citations, and to an extent usage, are not adequate measures of the impact or quality of published research (Seglen 1997; Aksnes 2006; Haustein and Larivière 2015) – this paper takes no position on the debate around the strength of usage or citations as a measure of impact, but recognizes they have been used as such and so is instead specifically interested in the relationship between citations and usage, as opposed to the validation of either as measures of impact.

With this in mind, we set out to further study the relationship between online article usage and citations using data from the world's largest, open access mega journal, *Scientific Reports* (Davis 2017). As *Scientific Reports* covers all areas of the natural and clinical sciences and it publishes a large volume of articles, it provides an excellent subject on which to carry out a large-scale study of the relationship between usage and citations, while also allowing us to look at additional breakdowns of this relationship (namely the influence of time and subject areas). These elements also mean that *Scientific Reports* is atypical when it comes to the usual structure and output of scholarly journals and so potentially the choice to use *Scientific Reports* may mean that the findings of this paper are not applicable to journals not structured in the same way or covering the same multidisciplinary output. However, this study analyses more than 6,000 *Scientific Reports* articles with the aim of understanding, through an observational analysis of historical data, the relationship between usage and citations – including their correlation, the influence of timeframe on both citations and usage, and the differences between subject areas.

# The data

## Initial data collection

To generate the dataset on which we carried out the analysis of citations and usage data, we extracted article information (Digital Object Identifiers and Publication Date) for all articles published in *Scientific Reports* between 2012 and 2014 from the journal's publishing platform. As *Scientific Reports* is a multidisciplinary journal and has a high rate of publication, focusing on this time period ensured we had a sizeable group of articles (7381) that had been published across an assessable period of time (2 years) and also covered a broad range of subject areas. *Scientific Reports* only publishes two types of article content, 'Research' and 'Amendments and Corrections'; as we were only interested in understanding the relationship between citations and usage for published research, we removed all 'Amendments and Corrections' from our record counts (n = 159), leaving us with just the 'Research' articles published in this period (n = 7222).

For each article, we extracted its citation counts for the first two years since publication and its usage statistics for the first year since publication. We decided to use these different timeframes as we know from a number of previous studies (e.g. Perneger 2004, Moed 2005, Brody et al. 2006) that there is a delay in the time it takes for articles to accrue citations, whereas usage data start being generated immediately. To extract each article's citation counts we utilized the API of the Scopus database (https://www.scopus.com/home.uri), which enabled us to collect monthly citation counts over two years for each article - giving us 24 monthly citation counts per article.

To extract each article's monthly usage data for the first year since publication we utilized WebTrends (https://www.webtrends.com), the web tracking system used by *Scientific Reports'* publisher (Nature Publishing Group) to monitor the journal's web activity. From this system we extracted three different monthly usage counts for each article (HTML views, PDF downloads and combined HTML and PDF usage), and we will explore these different usage counts in the next section. As our aim was to focus on the potential correlation (or at least association) between citations and usage, and so we only required cited articles, we removed all articles that had not been cited at the time of extraction (n = 373). We further removed those articles that were outliers in terms of usage (i.e. those that had more than 100,000 views), as there was a very small number of them (n = 6) and these had the potential to greatly skew the results of any analysis carried out. The resultant primary dataset was therefore made up of 6841 articles published in 2012–2014, which had been cited within two years of being published, and cover all the 68 different subject areas *Scientific Reports* lists on its website (https://www.nature.com/srep/browse-subjects).
Figure 1 contains a flow chart of the article selection process. The data were collected in January 2017 using Python scripts developed for this study, while the authors were employed by Springer Nature. No protocol of this study was pre-registered, and both the datasets and the scripts are available with permission from Springer Nature.

**Fig. 1** Flowchart of our article selection process

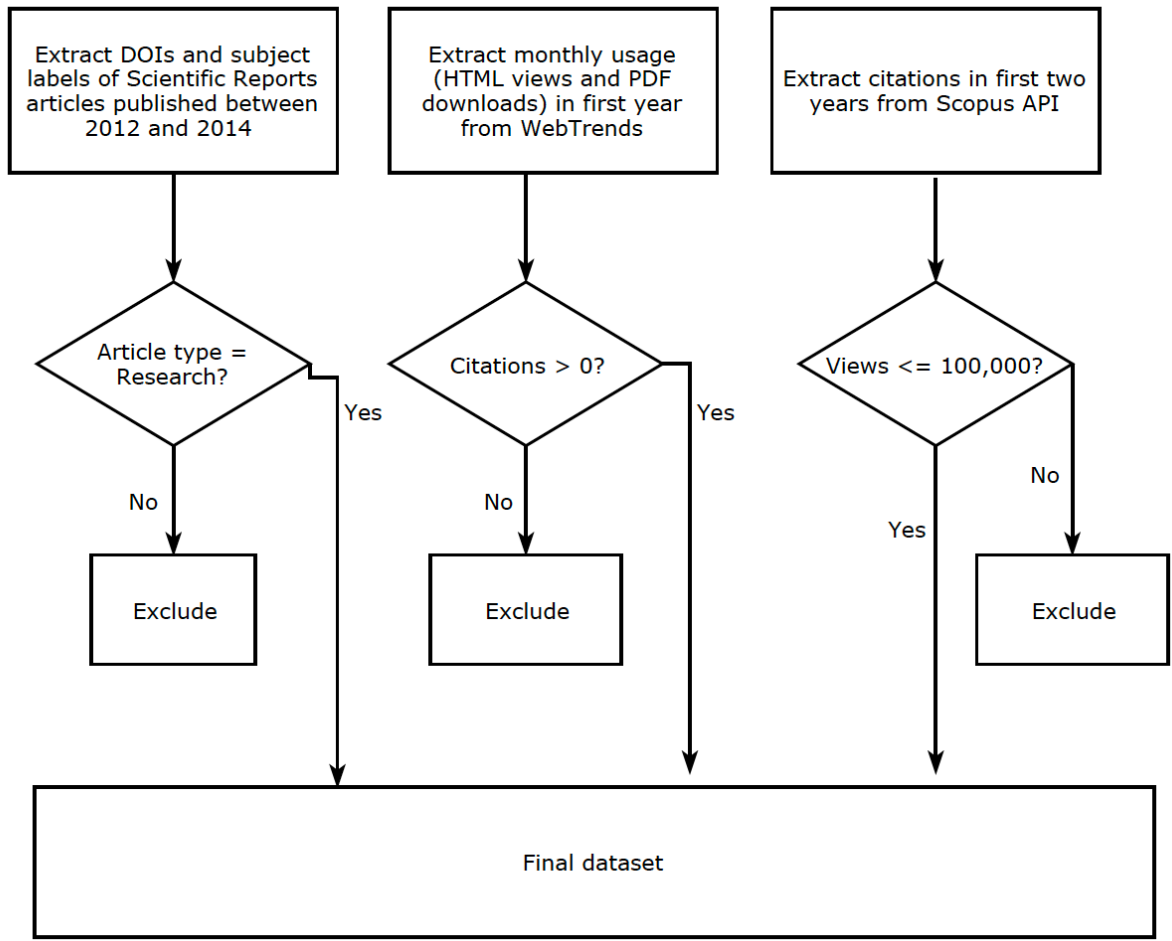

## Citation counts and definition of 'usage'

In our gathering of the data, we faced a potential difficulty in regard to "usage". While Scopus provided us with a single monthly citation count per article, meaning that our primary dataset had clear citation data for each article, WebTrends provided us with three different types of usage data per article – namely a count for HTML views per article, a count for PDF downloads per article and a count for combined HTML views and PDF downloads per article (hereafter referred to as 'HTML+PDF usage'). To identify which of these three different usage measurements was the most appropriate (and best suited) for our analysis, we first set about seeing if there was a difference in the way in which each of the usage counts were distributed and how these compared to the distribution of citations. Additionally, previous studies have focused specifically on PDF downloads' relationship with citations because, as one study puts it, "they measure at least the intention to use the downloaded material" (Gorraiz et al.2014). So, beyond our main aim of investigating the relationship between usage and citations, we were also tangentially interested in the question of whether such an assertion was correct, and if different types of usage measurements had different relationships with citations.

Therefore, we looked at the distribution of total citations over the first 24 months of an article's life, as well as the distribution of the totals for each of the different usage measurements (HTML views, PDF downloads and HTML+PDF usage) in the first 12 months after an article had been published.

**Fig. 2** Distribution of the articles' total citation counts over two years since publication

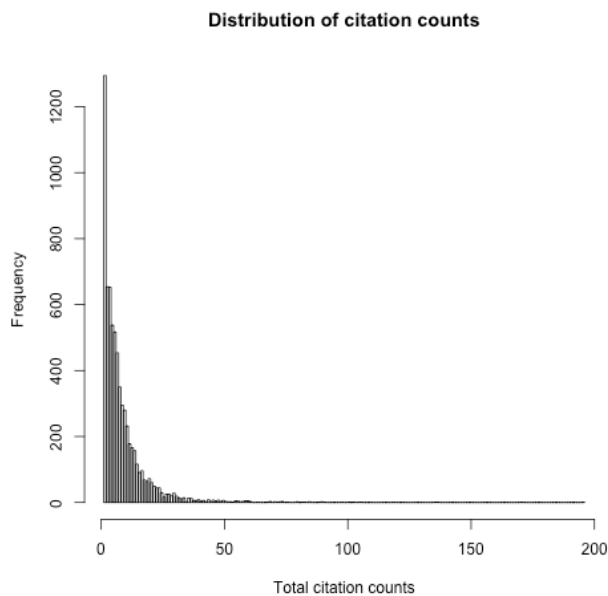

**Fig. 3** Distribution of total HTML+PDF usage counts over one year since publication

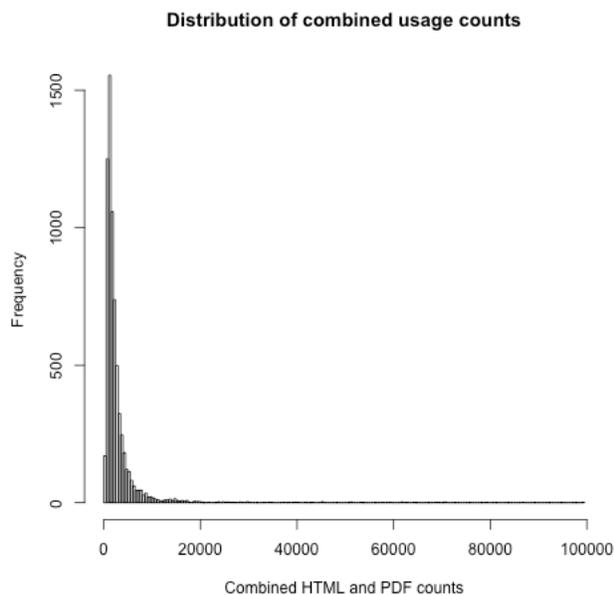

We found that the distributions of all elements (citations, HTML views, PDF downloads, HTML+PDF usage) were not normal. The skewed distributions for citations and HTML+PDF usage can be seen in *Figure 2* and *Figure 3*, with HTML views and PDF downloads' distributions being very similarly skewed, as can be seen in the specifics of each of these distributions detailed in the next paragraph.

The distribution of citations is very skewed, with total counts ranging from 1 to 196, a mean of 8.9, a median of 6, 90% of the articles having less than 19 citations and 70% having less than 10. The distribution of HTML+PDF usage is also very skewed, ranging from 207 to 99,365, with a mean of 2,746, a median of 1,686, 90% of the articles having less than 5,081 counts and 70% having less than 2,523. HTML views are similarly distributed ranging from 0 to 80,990, with a mean of 1,607, a median of 1,035, 90% of the articles having less than 2,843 counts and 70% having less than 1,501. The same holds for PDF downloads, which range from 0 to 83,510, with a mean of 1,139, a median of 548, 90% of the articles having less than 2,051 counts and 70% having less than 866.

In order to detect any difference between each of the measurements of usage, we decided to include all of them in our following analysis. Therefore, where relevant, our study will be broken into three different sets of analysis in regard to usage:
1. HTML views vs citations
2. PDF downloads vs citations
3. HTML+PDF usage vs citations

## Definition of subject area groups

To enable us to carry out an analysis of the relationship between citations and usage at the subject level, we also broke down our dataset into subject area groups using *Scientific Reports*' standardized structure for subject tagging. The journal's tagging system is based on author-selected keywords (hereafter referred to as 'subject area tags'), which are chosen at the point of submission and inserted into the article's XML at publication – these also appear on the article page online. There are hundreds of subject area tags available, but all of those contained in our dataset fall under four top-level subject areas: biological sciences, Earth & environmental sciences, health sciences, and physical sciences.

Articles can have as many subject area tags associated with them as the article's authors choose, as these tags are not curated by the journal's editorial staff. This presents a problem when aiming to assess the relationship between citations and usage at the subject level, because authors may not have chosen subject areas tags that are actually appropriate to the subject area of the article, due to human error or a misunderstanding of the tags. To minimize the effect of this, we only grouped articles based on the four top-level subject areas, using the reasoning that even if an author selected some irrelevant subject tags, the primary subject area(s) the article relates to would be captured in the top-level subject areas.

Even after we reduced the number of subject areas to the four top-level areas, articles still appeared in multiple subject areas. As there would be double counted articles if we simply carried out an analysis of each area, we looked at the distribution of every tagging combination across the four top-level subject areas (see *Table 1*).

| Row number | Biological sciences | Earth & Environmental sciences | Health sciences | Physical sciences | Number of articles |
| --- | --- | --- | --- | --- | --- |
| 1 | 0 | 0 | 0 | 1 | 2432 |
| 2 | 1 | 0 | 0 | 1 | 1342 |
| 3 | 1 | 0 | 1 | 0 | 856 |
| 4 | 1 | 0 | 0 | 0 | 850 |
| 5 | 1 | 0 | 1 | 1 | 379 |
| 6 | 1 | 1 | 0 | 0 | 294 |
| 7 | 0 | 1 | 0 | 1 | 284 |
| 8 | 1 | 1 | 0 | 1 | 189 |
| 9 | 0 | 0 | 1 | 0 | 106 |
| 10 | 0 | 1 | 0 | 0 | 36 |
| 11 | 0 | 0 | 1 | 1 | 27 |
| 12 | 1 | 1 | 1 | 0 | 19 |
| 13 | 1 | 1 | 1 | 1 | 17 |
| 14 | 0 | 1 | 1 | 0 | 7 |
| 15 | 0 | 1 | 1 | 1 | 3 |

*Table 1: Distribution of articles in each combination of top-level subject tags. In columns 2-5 "0" and "1" refer to the absence and presence of the subject tags, respectively.*

We identified 614 articles that covered three or more of the four top-level subject areas. These highly multidisciplinary articles (rows 5, 8, 12-14, in *Table 1*) were not distinct enough to include in a meaningful subject analysis and so we removed them from the sample.

Furthermore, our analysis of the remaining multidisciplinary articles found that there was a high degree of variation in the different subject tags to make any meaningful assumption that articles in these groups represented a distinct subject group (as opposed to simply 'multidisciplinary' research) and therefore we removed articles that contained any of the following tagging combinations:
- 'Biological sciences' and 'Physical sciences' (row 2 in *Table 1*, 1342 articles)
- 'Biological sciences' and 'Earth and environmental sciences' (row 6 in *Table 1*, 294 articles)
- 'Health sciences' and 'Physical sciences' (row 11 in *Table 1*, 27 articles)

We were then left with six groups of articles (rows 1, 3, 4, 7, 9 and 10 in *Table 1*); of these, the 36

articles that had only 'Earth and environmental sciences' tags (and no other subject area tags) or the 106 articles that had only 'health sciences' tags (and no other subject area tags) were comparatively too few (row 10 and row 9 in *Table 1*, respectively) to lead to any definite conclusions. Because the size of the remaining groups varied considerably (ranging between 36 articles and 2,432 articles), and the topics covered by groups of the top-level subject (namely 'biological sciences' and 'health sciences'; and 'Earth and environmental sciences' and 'physical sciences') were reasonably similar, we settled on combining the following six article groups into two distinct subject area groups:

- 'Bio-Health' class (1812 articles), which contains:
  - all articles with 'biological sciences' tags and without 'physical sciences' tags and without 'Earth and environment sciences' tags and without 'health sciences' tags (row 4 in *Table 1*, 850 articles)
  - all articles with 'health sciences' tags and without 'physical sciences' tags and without 'Earth and environment sciences' tags and without 'biological sciences' tags (row 9 in *Table 1*, 106 articles)
  - all articles with both 'biological sciences' and 'health sciences' tags and without 'physical sciences' tags and without 'Earth and environment sciences' tags (row 3 in *Table 1*, 856 articles)
- 'Phys-Earth' class (2752 articles), which contains:
  - all articles with 'physical sciences' tags and without 'biological sciences' tags and without 'health sciences' tags and without 'Earth and environment sciences' tags (row 1 in *Table 1*, 2432 articles)
  - all articles with 'Earth and environment sciences' tags and without 'biological sciences' tags and without 'health sciences' tags and without 'physical sciences' tags (row 10 in *Table 1*, 36 articles)
  - all articles with both 'physical sciences' and 'Earth and environmental sciences' tags and without 'biological sciences' tags and without 'health sciences' tags (row 7 in *Table 1*, 284 articles)

These final two subject area groups ('Bio-Health' and 'Phys-Earth') have no overlap with each other and so provide two distinct subject area groups on which solid conclusions can be drawn in regard to their impact on the relationship between usage and citations.

## Analysis

### Overlap of 'top cited' and 'top usage'

We began our analysis with the following question: are the top cited articles (in our dataset) also those with the top usage counts?

To find out the answer to this we identified the articles that were among the top 10% cited articles (those articles with 19 citation or more and which are hereafter referred to as 'Top cited') for the dataset as a whole and for each subject area group. We also identified those articles which were in the top 10% for article usage (broken down by each of our three usage measurement groups, hereafter each are referred to 'Top HTML views', 'Top PDF downloads' and 'Top HTML+PDF usage', or 'Top usage' collectively). We carried out an analysis of each of these subsets to calculate the overlap between 'Top cited' and 'Top usage' for all articles in the dataset, as well as specifically for the 'Bio-Health' and 'Phys-Earth' subject area groups.

|  | All articles (full dataset) | Bio-Health | Phys-Earth |
|---|---|---|---|
| **'Top cited' articles** | 637 | 178 | 259 |
| **'Top HTML views' articles** | 684 | 182 | 275 |
| **No. of overlapping 'Top cited' and 'Top HTML views' articles** | 258 | 65 | 117 |
| **Overlap between 'Top cited' and 'Top HTML views' (%)** | 40.5 [36.7,44.4] $\kappa$ = 0.33 [0.29,0.36] | 36.5 [29.5, 44.0] $\kappa$ = 0.29 [0.22,0.36] | 45.2 [39.0, 51.5] $\kappa$ = 0.38 [0.32,0.43] |
| **'Top PDF downloads' articles** | 683 | 182 | 276 |
| **No. of overlapping 'Top cited' and 'Top PDF downloads' articles** | 245 | 59 | 95 |
| **Overlap between 'Top cited' and 'Top PDF downloads' (%)** | 38.6 [34.7, 42.3] $\kappa$ = 0.30 [0.27,0.34] | 33.1 [26.4, 40.6] $\kappa$ = 0.25 [0.19,0.32] | 36.7 [30.9, 42.9] $\kappa$ = 0.29 [0.23,0.34] |
| **'Top HTML+PDF usage' articles** | 684 | 182 | 276 |
| **No. of overlapping 'Top cited' and 'Top HTML+PDF usage' articles** | 261 | 68 | 110 |
| **Overlap between 'Top cited' and 'Top HTML+PDF usage' (%)** | 41.0 [37.1, 44.9] $\kappa$ = 0.33 [0.29,0.37] | 38.2 [31.1, 45.8] $\kappa$ = 0.31 [0.24,0.38] | 42.5 [36.4, 48.8] $\kappa$ = 0.35 [0.29,0.40] |

*Table 2: Counts for 'Top cited' articles and 'Top' usage measurements (HTML views, PDF downloads, and HTML+PDF usage), including overlap between each group in terms of both number of articles and percentage values, followed by 95% confidence intervals, Cohen's Kappa, and Cohen's Kappa's confidence intervals.*

For this analysis we kept the definition of 'Top cited' articles for each of the groups ('All articles', 'Bio-Health' and 'Phys-Earth') the same (i.e. citation count should exceed 19) when comparing with each of our usage measurement groups, as this enabled us to have a clear understanding of the difference in effect across the three groups.

As can be seen in *Table 2*, the number of 'Top cited' articles identified for this overlap analysis was 637 in the 'All articles' group, 178 when looking at the 'Bio-Health' group and 259 for the 'Phys-Earth' group. The discrepancy in totals between the subject area groups and all articles is due to the fact that 200 of the 'Top cited' articles fell into subject area groups that were not included in our subject-level analysis.

Based on the number of articles in the overlap between the top cited and the top usage articles, according to each of the measures considered (HTML views, PDF downloads, and combined HTML+PDF usage), we calculated the proportion of these numbers with respect to the number of articles in the top cited group, and obtained 95% confidence intervals using the 1-sample proportions test with continuity correction in the statistical programming language R (R Core Team 2017). The results show that the confidence intervals overlap in all cases, leading us to the conclusion that there is no significant difference between the proportions in the different groups, both across the subject areas considered and the usage metrics analyzed.

Through this analysis we can see that, when using 'HTML+PDF usage' one could likely, in most cases, identify a closer relationship between citations and usage than when simply using 'HTML views' or 'PDF downloads' (although the difference between the different usage types is very small). While there is a noticeable overlap between the articles that are most used and those that are top cited, this overlap does not include all or even the majority of articles, and so citations and usage cannot be said to be interchangeable.

This led us to conduct a more in-depth analysis aimed to identify the relationship between these metrics, by looking at the correlation of total citations with total usage, as described in the next section.

## Correlations of Citations and Usage

As we found that there was a reasonable but not major overlap between the articles in our dataset that are most cited and those that have the top usage, we expanded the scope of our analysis to better understand the relationship between citations and usage. To do this we broadened our focus to look at the total numbers for citations and usage (as opposed to looking only at those articles that scored highest for these metrics), with the aim to identify the overall correlation between these groups. With our understanding of the different ways and time frames in which citations and usage metrics are accrued on articles, we calculated total counts as follows:

Usage = total number of usage counts (according to all three usage metrics) at the end of the first year after publication;
Citations = total number of citations at the end of the second year after publication.

Articles take some time to accrue citations (the mean month of first citation is 6.303), whereas usage metrics begin to accrue immediately after publication. Therefore, we decided to consider citations over the first two years since publication and usage counts over the first year since publication. This has two main benefits. First, it ensures that there is enough data in both distributions analyzed in the correlation (usage and citations) to enable robust correlation results. Second, by using this approach we also created the basis for understanding the role of time in the relationship between usage and citations. Starting from the selected timeframes (a year for usage and two years for citations), we went on to refine these in a more detailed analysis, reported in a later section of this article, 'Role of time in the impact of usage on citations'.

**Correlation tests on total usage and total citations**

To find the overall correlation between total usage and total citations, we ran a Spearman correlation test on the total number of citations at the end of the second year after publication and the total number of 'HTML+PDF usage' at the end of the first year since publication for each article. We also ran the same test using 'HTML views' and then again using 'PDF downloads'. The tests returned the results shown in *Table 3*, all statistically significant at the 0.05 level.

| Usage metric | Correlation coefficient |
|---|---|
| HTML+PDF usage | 0.49 [0.47,0.51] |
| HTML views | 0.47 [0.45,0.49] |
| PDF downloads | 0.49 [0.47,0.51] |

*Table 3: Spearman correlation coefficients (and their confidence intervals) between total usage at the end of the first year after publication and total citations at the end of the second year after publication for all articles, broken down by 'HTML+PDF usage', 'HTML views', and 'PDF downloads'. The results are statistically significant (α=0.05).*

The tests on the two subject area groups ('Bio-Health' and 'Phys-Earth'), produced the results shown in *Table 4* and *Table 5*, all statistically significant at the 0.05 level.

| Usage metric | Correlation coefficient |
|---|---|
| HTML+PDF usage | 0.43 [0.40,0.47] |
| HTML views | 0.43 [0.40,0.47] |
| PDF downloads | 0.53 [0.50,0.56] |

*Table 4: Spearman correlation coefficients (and their confidence intervals) between total usage at the end of the first year after publication and total citations at the end of the second year after publication for 'Bio-Health' articles, broken down by 'HTML+PDF usage', 'HTML views', and 'PDF*

*downloads'. The results are statistically significant (α=0.05).*

| Usage metric | Correlation coefficient |
|---|---|
| HTML+PDF usage | 0.53 [0.50,0.56] |
| HTML views | 0.53 [0.50,0.56] |
| PDF downloads | 0.53 [0.50,0.56] |

*Table 5: Spearman correlation coefficients (and their confidence intervals) between total usage at the end of the first year after publication and total citations at the end of the second year after publication for 'Phys-Earth' articles, broken down by HTML+PDF usage', 'HTML views', and 'PDF downloads'. The results are statistically significant (α=0.05).*

From *Tables 3-5* we can see that there is a statistically significant, moderate to strong correlation between usage and citations, if we interpret the coefficients following Cohen (1988)'s standard. We can also see that for 'Bio-Health' articles this correlation tends to be stronger when just looking at 'PDF downloads' of an article as opposed to 'HTML views'.

We see clearly that the correlation between usage and citations is markedly more pronounced in 'Phys-Earth' articles than in 'Bio-Health' articles (0.53 vs 0.43), and we hypothesize that this may be down to the fact that 'Bio-Health' articles tend to display a higher degree of variance in usage data than 'Phys-Earth' articles. This level of correlation means that in most cases it is safe to assume that when an article's usage (no matter its subject area) increases, so will its citations, albeit at a later date. But at which point in those first 12 months of usage can one start to comfortably make this assumption? To answer this, we moved on to assess the change in correlation between total citations in the first two years and total usage over the first 12 months after an article has been published.

**Correlations by month for all article groups**

To understand how the correlation between total usage and total citations changes over the first 12 months after an article is published, we ran another set of Spearman correlation tests. We ran a Spearman test between the total citation counts at the end of the second year and the cumulative 'HTML+PDF usage' at the end of each of the months from month 2 to month 12, we thus produced 33 test results, across the three groups of 'All articles', 'Bio-Health' and 'Phys-Earth'.

| Month since publication | All articles | Bio-Health | Phys-Earth |
|---|---|---|---|
| 2 | 0.39 [0.37,0.41] | 0.36 [0.32,0.40] | 0.42 [0.38,0.45] |
| 3 | 0.42 [0.39,0.44] | 0.38 [0.33,0.42] | 0.45 [0.41,0.47] |
| 4 | 0.44 [0.42,0.46] | 0.40 [0.35,0.44] | 0.46 [0.43,0.49] |
| 5 | 0.45 [0.43,0.47] | 0.41 [0.37,0.45] | 0.48 [0.45,0.51] |
| 6 | 0.46 [0.44,0.48] | 0.42 [0.38,0.46] | 0.49 [0.46,0.42] |
| 7 | 0.47 [0.46,0.49] | 0.43 [0.38,0.46] | 0.50 [0.46,0.53] |
| 8 | 0.48 [0.46,0.50] | 0.43 [0.39,0.50] | 0.51 [0.48,0.54] |
| 9 | 0.48 [0.46,0.50] | 0.43 [0.39,0.50] | 0.51 [0.49,0.55] |
| 10 | 0.49 [0.47,0.51] | 0.43 [0.39,0.47] | 0.52 [0.49,0.55] |
| 11 | 0.49 [0.47,0.51] | 0.43 [0.39,0.47] | 0.52 [0.49,0.55] |
| 12 | 0.49 [0.48,0.51] | 0.43 [0.39,0.47] | 0.53 [0.50,0.56] |

*Table 6: Spearman correlation coefficients (and their confidence intervals) of cumulative 'HTML+PDF usage' counts at each month from 2 to 12 and total citations at the end of the second year after publication, categorized into 'All articles', 'Bio-Health' and 'Phys-Earth'.*

As would be expected, due to the cumulative nature of the counts used in these correlation tests, the coefficients across all groups increase as the number of months increases, but it is notable that there is a period of increase across all groups which then levels off. For the 'All articles' group, the coefficient increases steadily over the first 7 months after publication and levels out momentarily at month 8 and 9, increases to 0.49 at month 10 and stays at this rate until the end of the 12-month period (as can be seen in *Figure 4*). From this levelling off at 10 months, at a coefficient value of 0.49, we can reasonably make the assertion that 10 months after publication it is possible to utilize usage to (at least partially) understand the level of citations an article will receive in comparison to other articles of the same age. As the correlation is only moderate (and strong for 'Phys-Earth' articles) according to Cohen (1988)'s standard, one cannot predict the number of citations at the end of the 2nd year after publication purely based on the usage accrued at 10 months; however, one could reasonably state that an article with higher usage at 10 months is likely to have more citations after 24 months.

**Fig. 4** Distribution of correlation coefficients between total citations and cumulative usage for months 2-12 after publication ('All articles' group)

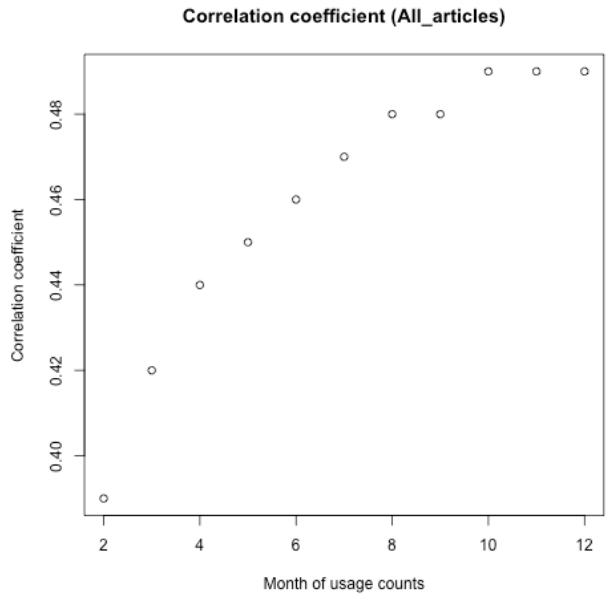

**Fig. 5** Distribution of correlation coefficients between total citations and cumulative usage for months 2-12 after publication ('Bio-Health' group)

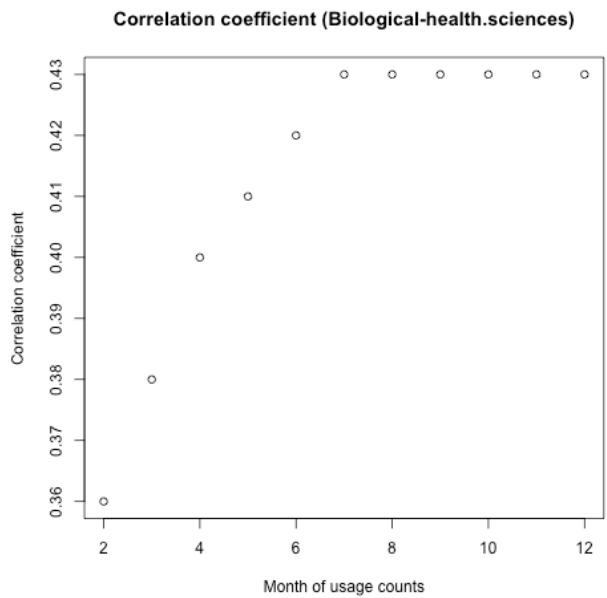

The increase of coefficients for 'Bio-Health' articles levels off much sooner than the other two groups (as seen in *Figure 5*). The correlation coefficient for this group of articles at the beginning of the time frame does not reach the same level as the other groups (so the correlation between usage and citations is weaker for these articles), but it does level off at an earlier time. After an article has been available for six months, the coefficient levels off at 0.43 and stays at this level until month 12. This means that some relationship between usage and citations for 'Bio-Health' articles can be stated by month six, but as the coefficient is lower for this group (and so the correlation moderate), the confidence one can have in making these assumptions is lower than with the other groups of articles.

**Fig. 6** Distribution of correlation coefficients between total citations and cumulative usage for months 2-12 after publication ('Phys-Earth' group)

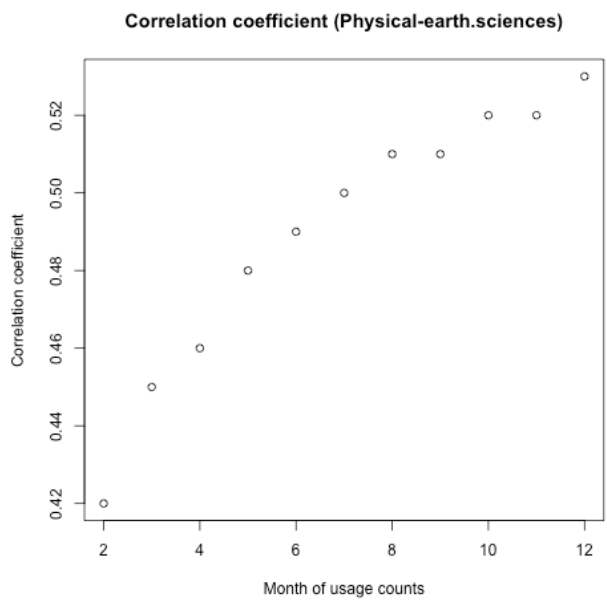

Finally, for 'Phys-Earth' articles we see a similar increase in coefficient levels as in the other groups, albeit starting from a higher value (0.42 at month 2). As for the 'All Articles' group, the coefficients of the 'Phys-Earth' articles level off between month 8 and month 9 and then again between month 10 and month 11, with a slight increase at month 12, ending with a coefficient of 0.53. This final coefficient is considerably higher than the one of the 'Bio-Health' group, which may suggest that we need to look at usage counts beyond the first year before the rate of increase of this correlation levels off completely. This higher coefficient at month 12 enables us to make a more comfortable assertion that the rate of usage in the first year for an article can be related to the level of citations that the article will have two years after publication.

Based on these correlation tests of citation and usage (for all articles and the different usage and subject area groups within our dataset), we can confidently say that there is a moderate correlation between the level of usage in the first 12 months since publication and the number of citations 24 months after publication. Moreover, general assumptions can be drawn ten months after publication about the level of citations an article will have after two years (in relation to similarly-aged articles); this can likely start by month 6-7 for 'Bio-Health' articles, but can be drawn with more confidence for 'Phys-Earth' articles.

So far the two approaches we have taken to explore the relationship between citations and usage (comparing 'top cited' and 'top usage' articles, and looking at the correlation between total citations and total usage) have shown that there is a relationship between these two metrics, but this relationship has a time aspect that varies depending on an article's subject areas and we can only be moderately confident in the effect one will have on the other (primarily the effect usage will have on citations, as usage is accrued earlier than citations). Therefore, this paper will close by exploring if we can infer a more precise relationship between the two metrics by looking closer at the role time plays in their accrual.

## Role of time in the impact of usage on citations

In the final section of this paper we will explore the role time plays in regard to the relationship between usage and citations. To do this we will look at the following questions:
- How is usage accrued in the first 12 months?
- How does usage relate to the time of first citation of an article?
- How does the time of first citation of an article relate to the total number of citations at the end of the first 24 months since publication?

**Usage levels and time of first citation**

We calculated the mean month of first citation for all articles, which is 6.30 (6.89 for 'Bio-Health' and 5.99 for 'Phys-Earth'), and the median month of first citation, which is 5 (6 for 'Bio-Health' and 5 for 'Phys-Earth'), respectively. To better understand the role time plays in the relationship between usage and citations, and as the usage metrics used in our analyses are cumulative, we decided to reduce the timeframe for analyzing how usage related to time of first citation to look at just the usage accrued in first six months. By narrowing the time window in this way, we were able to reduce the potential noise created by looking at the full 12 months – in particular the possible impact of a 'double effect' resulting from an increase in usage following an article's additional exposure from being cited.

We defined 'top used' articles in each group as those articles that appear in the top 10% of all articles according to their usage counts in the first six months. We then compared this group with all other articles with respect to the month of their first citation. As shown in Table 7, the mean month of first citation for top used articles is 5.25 and the median is 4, while the mean month of first citation for all other articles is 6.42 and the median is 5. There seems to be evidence for the fact that when an article is used more in the first six months, it is more likely to be also cited earlier. The mean month of first citation for 'Phys-Earth' articles is 5.99, and for highly downloaded articles it is 5.29. The mean month of first citation for 'Bio-Health' articles is 6.89, and for highly downloaded articles it is 5.95. So, Physical-Earth articles seem to have more downloads and are cited sooner than 'Bio-Health' papers.

|  | Article group | All articles | 'Bio-Health' articles | 'Phys-Earth' articles |
|---|---|---|---|---|
| All article usage | Mean month of first citation | 6.30 | 6.89 | 5.99 |
|  | Median month of first citation | 5 | 6 | 5 |
|  | Standard deviation of first citation | 4.96 | 5.28 | 4.71 |

|  | Interquartile range of first citation | 6 | 7 | 5 |
|---|---|---|---|---|
| 'top used' articles in first 6 months | Mean month of first citation | 5.25 | 5.76 | 5.10 |
|  | Median month of first citation | 4 | 4 | 4 |
|  | Standard deviation of first citation | 4.36 | 4.82 | 4.12 |
|  | Interquartile range of first citation | 5 | 6 | 4 |
| 'non-top used' articles in first 6 months | Mean month of first citation | 6.42 | 7.02 | 6.09 |
|  | Median month of first citation | 5 | 6 | 5 |

*Table 7: Mean, median, standard deviation, and interquartile range month of first citation for articles in the 'All articles', 'Bio-Health' and 'Phys-Earth' article groups, for all article usage and for usage in the first six months.*

**Fig. 7** Box-and-whiskers plot of month of first citation for 'top used' articles, and all other articles

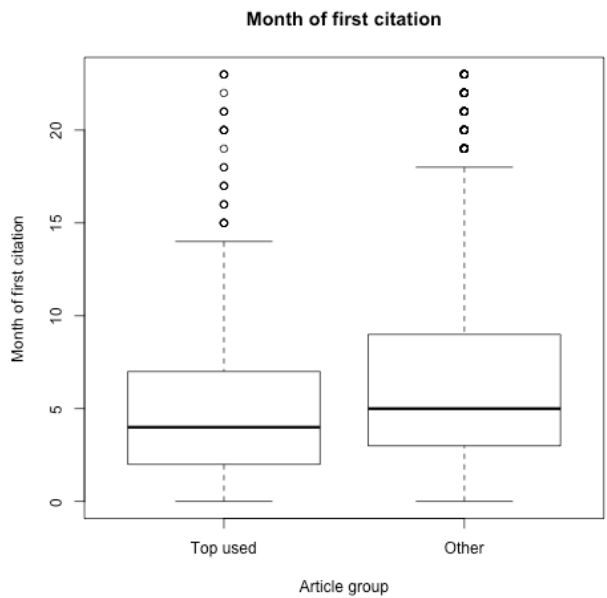

The box-and-whiskers plot for the distribution of the month of first citation for the two groups of articles contained in *Figure 7* (which is very similar to those for the 'Bio-Health' and 'Phys-Earth' groups) shows that the articles which were 'highly used' in the first 6 months seem to be more likely to have their first citation occur earlier. In order to test this statistically, we performed a one-sided Wilcoxon rank sum test with continuity correction between the two distributions. The results (W=1811500, p < 0.0001) show a statistically significant difference, with the top used articles having a lower median month of first citation compared to all other articles (the pseudo-median of the difference between the two is estimated to be -0.99). Therefore, it can be reasonably stated that if an article is 'highly used' in the first 6 months after publication, then it is more likely to be cited earlier than an article that is not used as much in the first 6 months. Very similar results hold for the 'Bio-Health' (W = 127180, p = 0.0007, estimate of pseudo-median of the difference = -1.00) and 'Phys-Earth' articles (W = 300860, p = 0.0005, estimate of pseudo-median of the difference = -0.99).

The mean month of first citation for 'Phys-Earth' articles is 5.99, and for highly downloaded articles it is 5.29. The mean month of first citation for 'Bio-Health' articles is 6.89, and for highly downloaded articles it is 5.95. So, Physical-Earth articles have more downloads and are cited sooner than 'Bio-Health' articles.

We can conclude then that if an article is 'highly used' in the first 6 months then it is more likely to be cited earlier, but what effect does being cited earlier have on overall citations in the first 24 months?

**Time of first citation and total citations**

To find the effect that being an early cited article has on an article's total citations after 24 months, we defined 'early cited' articles as those articles whose first citation occurs before the median month of first citation (which is 5 for all articles, 6 for 'Bio-Health' and 5 for 'Phys-Earth').

| Article group | Mean total number of citations for all articles | Median total number of citations for all articles | Mean total number of citations for 'early cited' articles | Median total number of citations for 'early cited' articles | Mean total number of citations for 'not-early cited' articles | Median total number of citations-for 'not early cited' articles |
|---|---|---|---|---|---|---|
| All articles | 8.91 | 6 | 12.1 | 9 | 6.46 | 4 |
| 'Bio-Health' articles | 6.71 | 5 | 8.98 | 7 | 4.68 | 4 |
| 'Phys-Earth' articles | 10.63 | 7 | 14.17 | 10 | 7.73 | 5 |

*Table 8: Mean and median for total citations of articles in the 'All articles', 'Bio-Health' and 'Phys-Earth' article groups*

We performed a Wilcoxon test on the distribution of total citations of early cited articles and the distribution of total citations of all other articles. The results (W=8071200, p < 0.0001) are statistically significant and show that the early cited articles have a median of total citations that is higher than that of all other articles (the pseudo-median of the difference between the two is estimated to be 4.00). The results are very similar in the case of 'Bio-Health' (W = 590660, p <

0.0001, estimate of pseudo-median of the difference = 2.99) and 'Phys-Earth' articles (W = 1303400, p < 0.0001, estimate of pseudo-median of the difference = 4.00). Through this analysis we found that, across the board, when an article is cited earlier it is cited more over the first 24 months after publication (as can be seen from *Table 8*). Very similar results were obtained for the 'Bio-Health' and 'Phys-Earth' group and so we can safely assume that the results derived from this analysis for all groups are suitably robust.

In the 'All articles' group we found that the mean number of total citations of all 6841 articles is 8.9 citations after 24 months since publication, with a median of 6 (across a range of 1-196). However, of the 2964 'early cited' articles in this group, the mean number of total citations is 12.1 with a median of 9. The difference between the mean number of total citations of all articles and 'early cited' articles in this group represents an average uplift of 36%. Therefore, we can comfortably assert that when an article is cited early it is much more likely to have higher total citations 24 months after publication.

Likewise, in the 'Bio-Health' group we found that the mean number of total citations for all 1812 articles in this group is 6.7, with a median of 5 (and a range of [1,74]), while the mean number of total citations for the 856 'early cited' articles is 9.0 with a median of 7. For the 'Bio-Health' group, this represents a 34% uplift in the average number of citations for 'early cited' articles. So once again we find that those articles that are cited early have a greater number of total cites at the end of their first two years of publication. Finally, (perhaps unsurprisingly) we found a very similar story for the 2752 'Phys-Earth' articles. In this group the overall mean total citation is 10.6 with a median of 7 and a range of [1,196], whereas the mean number of total citations for the 1240 'early cited' articles increases to 14.1 and the median is 10, representing an average uplift in total citations of 33%.

Based on these results we can confidently state that if an article is cited early then it is more likely to be cited more overall (after two years) and it is likely to have a citation uplift of approximately 30% over those articles that are not 'early cited'

### Inferring the impact of usage on total citations

Throughout this section we have looked at the role time plays in the relationship between usage and citations, both from the point of view of how usage impacts when an article is first cited, as well as how the time of first citation of an article relates to the total number of cites that article has after 24 months.

At a simplified level the analysis in this section has shown that when an article is downloaded more in the first six months it is much more likely to be cited earlier and when an article is cited earlier it is much more likely to be cited more in total. We can therefore infer from this that when an article is downloaded more in the first six months it is more likely to be cited more overall within its first two years.

# Discussion and conclusion

Throughout this paper we have aimed to understand the relationship between usage and citations in the first two years of a published article's life. In exploring this relationship, we have looked at the correlation between the citations and usage data in articles published in a multidisciplinary, open access mega-journal through the lenses of types of usage, subject areas and the role of time. It should be noted that a correlation does not imply a causal relationship between the variables analyzed, and therefore we do not state causality in our results.

**Types of usage**

In relation to types of usage, this paper concludes that for this type of analysis one can very reasonably use HTML+PDF usage as the best proxy for article usage. While there is some variation between this metric and either HTML views or PDF downloads on their own, we have shown that it is more reliable to use HTML+PDF usage to identify a relationship between 'top cited' and 'top usage' articles and the difference between the three metrics is mostly negligible when looking at correlations between usage and citations. Therefore, this paper proposes that research looking at the usage of articles and its relationship with citations can successfully and reliably use the HTML+PDF usage metric for its analysis.

**Subject areas**

From a subject area perspective, we have found that for this dataset there appears to be a clear difference in the relationship between citations and usage when looking at different subject areas. While there is no statistically significant difference between the subject groups in the overlap between 'top cited' and 'top usage' articles, there is a noticeably higher correlation between the total usage and total citations of articles in the 'Phys-Earth' group when compared to the 'Bio-Health' group, and both showed a moderate to strong correlation (0.53 and 0.43, respectively), according to Cohen (1988)'s standard. Finally, we found that although 'Phys-Earth' articles are always more likely to be cited earlier on average than 'Bio-Health' articles and cited more on average in total after 24 months, 'Bio-Health' articles seem to see a greater benefit from being 'highly used', as well as from being 'early cited'.

**The role of time**

To better understand the role time plays in the relationship between citations and usage, we devised a variety of experiments. First, we looked at the role that being highly used within the first six months plays on the time when citations accrued and found that the more highly used an article is, the more likely it is to be cited earlier than average. We then moved on to analyze the impact of early citations on the overall number of citations an article will accrue in the 24-month period after publication: here we found a clear uplift (of ~30% on average) in overall citations for those articles that were cited early. The average cited half-life of papers in many of the subject areas covered by *Scientific Reports* is likely to be greater than 2 years and so it would potentially be of value to redo this study using longer timeframes, as long as other variables could be controlled in some way.

In this paper we have examined a number of aspects in the relationship between usage and citations. There is a greater amount of analysis and research that the results of this paper points towards: primarily, understanding more clearly the relationship between time and citations, and better understanding what role being a mega-journal plays in the results we have found and in the conclusions we have drawn. It should also be reiterated that *Scientific Reports* is somewhat atypical when it comes to the structure and output of scholarly journals more generally and so the results we have identified may not be reproduced if, for example, this study was replicated with a collection of different journals covering the same subject areas, etc. as we studied. We intend to explore these elements in future research projects and hope that the findings we have presented here are used as a basis for experimentation and analysis by others.

## Authors' contributions

BMcG collected and processed the data, and conducted the statistical analyses. BMcG was the major contributor to the Analysis and Methods sections. MA was the major contributor in writing the Introduction, Discussion and Conclusions sections and in interpreting the results. Both authors designed the study, read and approved the final manuscript.

## Acknowledgements

This work was supported by The Alan Turing Institute under the EPSRC grant EP/N510129/1.